\input harvmac
\noblackbox

\def\bfone{\relax{\rm 1\kern-.35em 1}}
\def\inbar{\vrule height1.5ex width.4pt depth0pt}

\def\IC{\relax\,\hbox{$\inbar\kern-.3em{\rm C}$}}
\def\ID{\relax{\rm I\kern-.18em D}}
\def\IF{\relax{\rm I\kern-.18em F}}
\def\IH{\relax{\rm I\kern-.18em H}}
\def\II{\relax{\rm I\kern-.17em I}}
\def\IN{\relax{\rm I\kern-.18em N}}
\def\IP{\relax{\rm I\kern-.18em P}}
\def\IQ{\relax\,\hbox{$\inbar\kern-.3em{\rm Q}$}}
\def\us#1{\underline{#1}}
\def\IR{\relax{\rm I\kern-.18em R}}
\font\cmss=cmss10 \font\cmsss=cmss10 at 7pt
\def\ZZ{\relax\ifmmode\mathchoice
{\hbox{\cmss Z\kern-.4em Z}}{\hbox{\cmss Z\kern-.4em Z}}
{\lower.9pt\hbox{\cmsss Z\kern-.4em Z}}
{\lower1.2pt\hbox{\cmsss Z\kern-.4em Z}}\else{\cmss Z\kern-.4em
Z}\fi}
\def\nup#1({Nucl.\ Phys.\ $\us {B#1}$\ (}
\def\plt#1({Phys.\ Lett.\ $\us  {B#1}$\ (}
\def\cmp#1({Comm.\ Math.\ Phys.\ $\us  {#1}$\ (}
\def\prp#1({Phys.\ Rep.\ $\us  {#1}$\ (}
\def\prl#1({Phys.\ Rev.\ Lett.\ $\us  {#1}$\ (}
\def\prv#1({Phys.\ Rev.\ $\us  {#1}$\ (}
\def\mpl#1({Mod.\ Phys.\ Let.\ $\us  {A#1}$\ (}
\def\ijmp#1({Int.\ J.\ Mod.\ Phys.\ $\us{A#1}$\ (}
\def\jag#1({Jour.\ Alg.\ Geom.\ $\us {#1}$\ (}
\def\tit#1|{{\it #1},\ }

\def\Coe#1.#2.{{#1\over #2}}
\def\coeff#1#2{\relax{\textstyle {#1 \over #2}}\displaystyle}
\def\coe#1.#2.{\relax{\textstyle {#1 \over #2}}\displaystyle}
\def\half{{1 \over 2}}

\def\del{\partial}

\def\br{\hfill\break}

%
%
\lref\MNW{J.~Minahan, D.~Nemeschansky and
N.P.~Warner, {\it Investigating the BPS Spectrum of Non-Critical 
$E_n$ Strings,} USC-97/006, NSF-ITP-97-055, hep-th/9705237.}
\lref\LMW{W.~Lerche, P.~Mayr and
N.P.~Warner, {\it Non-Critical Strings, Del Pezzo Singularities
and Seiberg-Witten Curves,} CERN-TH/96-326,USC-96/026,
hep-th/9612085.}
\lref\KMV{A.\ Klemm, P.\ Mayr and C.\ Vafa,
{\it BPS states of exceptional non-critical strings,}
CERN-TH-96-184, hep-th/9607139.}
\lref\KLMVW{A.\ Klemm, W.\ Lerche, P.\ Mayr, C.\ Vafa and
N.P.\ Warner,\nup{477} (1996) 746, hep-th/9604034.}
\lref\BCOV{M.\ Bershadsky, S. \ Ceccotti, H.\ Ooguri and C. \ Vafa,
\cmp{165} (1994) 311, hep-th/9309140.}
\lref\MandFWit{E.~Witten, \nup{471} (1996) 195,
hep-th/9603150.}
\lref\SW{N.\ Seiberg and E.\ Witten, \nup426(1994) 19,
hep-th/9407087; \nup431(1994) 484, hep-th/9408099.}
\lref\GSW{M.~Green, J.~Schwarz and E.~Witten, {\it Superstring Theory: 1},
Cambridge University Press, 1987.}
\lref\GMS{O.\ Ganor, D.\ Morrison and N.\ Seiberg,
\nup{487} (1997) 93,  hep-th/9610251.}
\lref\EnNCS{O.\ Ganor and A.\ Hanany, \nup{474} (1996) 122,
hep-th/9602120; \br
N.\ Seiberg and E.\ Witten, \nup{471} (1996) 121,
hep-th/9603003; \br
M.~Duff, H.~Lu and C.N.~Pope, \plt{378} (1996) 101,
hep-th/9603037; \br
M.R.~Douglas, S.~Katz, C.~Vafa, {\it
Small Instantons, del Pezzo Surfaces and Type I' theory,}
hep-th/9609071; \br
E.~Witten, \mpl{11} (1996) 2649, hep-th/9609159.}
\lref\DMCV{D.R.~Morrison and C.~Vafa, \nup{473} (1996) 74, hep-th/9602114;
\nup{476} (1996) 437, hep-th/9603161.}
\lref\MandFWit{E.~Witten, \nup{471} (1996) 195,
hep-th/9603150.}
\lref\Ganor{O. Ganor, \nup{479} (1996) 197, hep-th/9607020.}
%
%
\Title{\vbox{
\hbox{USC-97/009}
\hbox{NSF-ITP-97-091}
\hbox{\tt hep-th/9707149}
}}{\vbox{\centerline{\hbox{Partition Functions for BPS
States}}
\vskip 8 pt
\centerline{ \hbox{of the Non-Critical $E_8$ String}}}}
\centerline{J.A.~Minahan, D.~Nemeschansky}
\bigskip
\centerline{{\it Physics Department, U.S.C.}}
\centerline{{\it University Park, Los Angeles, CA 90089}}
\bigskip
\centerline{and}
\bigskip
\centerline{N.P.~Warner}
\bigskip
\centerline{{\it Institute for Theoretical Physics}}
\centerline{{\it University of California, Santa Barbara,
 CA 93106-4030 \footnote{*}{On leave from Physics Department,
U.S.C., University Park, Los Angeles, CA 90089}}}
\bigskip

\vskip .3in

We consider the BPS states of the $E_8$ non-critical string 
wound around one of the circles of a toroidal compactification
to four dimensions.  These states are indexed by their momenta
and winding numbers.  We find explicit expressions, $G_n$, for 
the momentum partition functions for the states with winding 
number $n$.  The $G_n$ are given in terms of modular forms.
We give a simple algorithm for generating the $G_n$, and
we show that they satisfy a recurrence relation that is
reminiscent of the holomorphic anomaly equations of
Kodaira-Spencer theory.

\vskip .3in


\Date{\sl {July, 1997}}

%
\parskip=4pt plus 15pt minus 1pt
\baselineskip=15pt plus 2pt minus 1pt
%
\newsec{Introduction}

One of the  surprises in the recent past is the existence of 
six-dimensional,
$E_8$ non-critical strings in heterotic string theory \EnNCS.  
These occur when a heterotic string is compactified on a $K3$ 
manifold with a small $E_8$ instanton.  As the instanton shrinks
to zero size, the tension of a non-critical string vanishes.   
One of the intriguing features of such strings is that
they are consistent string theories that are apparently decoupled 
from gravity.

If one compactifies the six-dimensional theory on a circle, then 
one finds BPS states that correspond to momentum and winding states 
of the non-critical string around the compactified dimension 
\refs{\Ganor,\KMV}.  
After a second compactification on another circle, we are left with 
an $N=2$ supersymmetric theory in four dimensions, where the
pre-potential has an instanton expansion whose coefficients basically 
count the net number of BPS states of the five-dimensional theory.

The six-dimensional heterotic theory is dual to an $F$-theory 
compactification  on a Calabi-Yau manifold that admits an 
elliptic fibration.  The tensionless string limit is reached when 
a del Pezzo $4$-cycle shrinks to zero size \refs{\DMCV,\MandFWit}.  
In \LMW\ it was shown how to describe the pre-potential of
the non-critical string as a function of the string tension
in terms of a Seiberg-Witten theory.  In \MNW\ this was extended 
to incorporate other physical moduli.

In this paper, we investigate this instanton expansion as a function
of two moduli: $t_S$ and $t_E$.  The former is a product of the
string tension and 
a compactification radius, and so indexes the
winding states around a circle.  The parameter $t_E$ is a background 
geometric factor that indexes the momentum states on the same circle.  
As an expansion in $t_S$, we find that each term of the pre-potential
is an almost 
modular function in $t_E$.  We  explicitly compute the first few 
functions in the expansion.  We then show that these functions 
satisfy a ``modular anomaly'' recurrence relation
which is similar to the non-holomorphic recurrence relation
of Kodaira-Spencer theory \BCOV.  Using the recurrence relation we
show how to generate all  functions in the instanton expansion.
We also have enough information about the structure of  these functions
to compute the BPS degeneracies in the asymptotic limit.

In section 2 we review some relevant facts from \refs{\KMV,\GMS,\LMW}
 and \MNW.
In section 3 we discuss the instanton expansion and its relation to the
counting of BPS states.  In section 4 we discuss and then prove the
recurrence relation.  We also show that the recurrence relation
is very useful in computing the entire instanton expansion.  In
section 5 we compute the asymptotic BPS state degeneracy.  

\newsec{The quantum effective action}

\subsec{The IIA description of the non-critical $E_n$ string}

We first briefly review some of the key elements of 
\refs{\KMV,\LMW}.
One considers a compactification of the IIA theory on  an 
elliptically fibered Calabi-Yau $3$-fold in 
which a $4$-cycle is collapsing.   The magnetic non-critical string 
may be thought of as coming from a five-brane wrapping this cycle,
and the electric excitations of the string come from
membranes wrapping the $2$-cycles within the $4$-cycle.  
One obtains the $E_n$ non-critical string if the collapsing
$4$-cycle is a del Pezzo surface, $B_n$, obtained by 
blowing-up $n$ points in $\IC \IP_2$\refs{\DMCV,\MandFWit}.  

In \KMV\ the foregoing is realized explicitly by using a 
Calabi-Yau $3$-fold, ${\bf X_{F_1}}$, 
that is an elliptic fibration over the Hirzebruch surface
${\bf F_1}$.  Since ${\bf X_{F_1}}$ is also a $K3$ fibration,
this compactification of the IIA theory is dual to the heterotic 
string compactified on $K3 \times T_2$.  Specifically, the 
corresponding heterotic string compactification has an 
$E_8 \times E_8$ instanton embedding with
$n_1 = 11, n_2 =13$.  The manifold  ${\bf X_{F_1}}$ has
three K\"ahler moduli, $t_D, t_E$ and $t_F$, corresponding to the
volumes of the base of ${\bf F_1}$, the elliptic fiber of 
${\bf X_{F_1}}$, and the fiber of ${\bf F_1}$ respectively.
The modulus, $t_D$, is also the scale of the canonical
divisor of the del Pezzo surface, and the elliptic fiber
of the del Pezzo is that of  ${\bf X_{F_1}}$.  At $t_D = 0$
only a $2$-cycle collapses, but for the whole del Pezzo to vanish
one must pass through a flop transition ($Im(t_D) < 0$) to a point where 
$t_D + t_E = 0$.   

Since $t_D$ is the scale of the base of the $K3$ fibration, it
follows that it must be related to the heterotic dilaton.
The other two moduli, $t_E$ and $t_F$, are related to the
moduli, $T = B + i R_5 R_6$ and $U = e^{i\alpha} R_5/R_6$ of
the torus in the heterotic compactification.  In \refs{\GMS,\LMW} it was 
shown 
that $S,T$ and $U$ are related to the complexified K\"ahler moduli
according to $t_E = U, t_F = T - U$ and $t_D = S + a T + b U$
for some undetermined constants $a,b$.  The point 
where $t_D$ vanishes corresponds to an
$SU(2)$ gauge symmetry enhancement at strong coupling in 
the heterotic string.

It is convenient to introduce a reparametrization
of the moduli, replacing $t_D$ and $t_F$ by 
$t_S \equiv t_D + t_E$ and $t_F' = t_F + t_B$.
(Note that both these new moduli are linear in the
heterotic dilaton.)  From the point of view of the non-critical 
string, the  interesting moduli are $t_E$ and $t_S$.
Of the three moduli, only $t_F'$ involves the tension
of the fundamental heterotic string \LMW, and so to decouple
the BPS states of fundamental heterotic string, we will
consider the limit in which $t_F' \to \infty$.
The parameter $t_S$ determines the scale of the $4$-cycle
and hence the tension of the non-critical string.  In terms of
the electric BPS excitations of the non-critical string
(membranes wrapping rational curves), the degrees $d_D$ and
$d_E$ of the rational curve represent the winding 
number and momentum respectively of the non-critical
string around one of the circles of the torus \KMV.

\subsec{Consistent truncations}

It is fairly evident that all the physics of the non-critical
string should be captured by the structure of the vanishing del
Pezzo surface, and  not so much by the details of the geometry of 
the $3$-fold in which the del Pezzo surface lives.
In the nearly tensionless limit, one should be able to
decouple the non-critical
string from ancilliary degrees of freedom like gravity.
This decoupling should be accomplished by finding
a way to abstract the del Pezzo surface from the space in which 
it is buried.  It was one of key ideas in \LMW\  that one 
can see how to do this in a consistent manner by passing to a closed 
sub-monodromy problem.  That is, one identifies a basis of cycles
that close (decouple from all the others) under monodromies on
a special subset of moduli.  Two such closed sub-monodromy
problems were identified in \LMW, and both of them
were on the other side of the flop transition ($Im(t_D) < 0$). 
One finds that in the basis where the scalar fields
in the vector multiplets are given by the parameters
introduced above, $t_E$, $t_S = t_E + t_D$ and
$t_F' = t_F + t_B$, three of the periods depend classically
upon $t_S$ alone, and one further period is $t_E$ itself.  

It was further proposed in \LMW\ that such closed
sub-monodromy problems could be modeled
by considering a compactification of the IIB string on
a non-compact Calabi-Yau $3$-fold based upon the
vanishing del Pezzo surface.  This non-compact Calabi-Yau
manifold would thus characterize the non-critical string
decoupled from the rest of the original string theory.
This proposal closely parallels
the way in which the quantum effective action
of gauge theories (decoupled from the rest of
string theory) can be obtained from ALE fibrations \KLMVW.
This idea was implemented and tested in \LMW, and further
tested in \MNW, for the three  period sub-monodromy problem 
that involves $t_S$ alone.  Our purpose here is to generalize 
this to include $t_E$.   

In a Calabi-Yau compactification of the IIB theory, 
the electric and magnetic BPS states are on the same footing
-- they come from wrapping $3$-branes around $A$ or $B$-cycles
of the Calabi-Yau manifold.  As in \refs{\KLMVW,\LMW},  
one can re-cast this wrapping of $3$-branes in terms of 
wrapping a non-critical string around a Riemann surface.
This is done by decomposing the $3$-cycles into 
$2$-cycles fibered over some curve in a base.  The holomorhpic 
$3$-form can then be integrated to yield some form of Seiberg-Witten 
differential.  Thus the original pre-potential can be
obtained from integrals of a meromorphic differential
on a Riemann surface.  For the non-critical string, such a 
Seiberg-Witten effective action encodes all the counting
of rational curves in the del Pezzo surface \refs{\LMW,\MNW}.
In \refs{\LMW,\MNW} this was done for sub-monodromy
problem involving $t_S$ alone, which yields information
about  the BPS states with
$d_D = d_E$.  These states were focussed on because they
were, in a sense, the most stringy: they are the states that become
massless when the non-critical string becomes tensionless.
The reduction of the problem from a $3$-fold to a torus was
also critical to \MNW\  in that it enabled the simple
generalization to include Wilson lines, and led to
a conjecture as to how to include $t_E$.

The starting point of \refs{\LMW, \MNW} was to consider the
IIB string theory compactified on the on the non-compact
Calabi-Yau $3$-fold defined by
\eqn\EeightCY{w^2 ~=~ z_1^3 ~+~ z_2^6 ~+~ z_3^6  ~-~
{1 \over z_4^6} ~-~ \psi\, w  z_1 z_2 z_3 z_4\ .}
The modulus $\psi$ determines the string tension.
As described in \LMW, the holomorphic $3$-form on this manifold 
has three periods corresponding to terms in the BPS mass formula.
With suitable normalization, the period integrals
can be identified with $1, t_S$ and $\del {\cal F}/\del t_S$.
This gives the truncation of the non-critical string to
the sector in which $t_S$ is the only parameter, and in
which the BPS states have $d_E = d_D$. 

To get the states of the string with independent
winding number and momentum ($d_D$ and $d_E$) we
want the slightly less stringent truncation described in
\LMW: one in which  the period integrals are 
$1, t_S, t_E$ and $\del {\cal F}/\del t_S$.  The 
relevant manifold was proposed in \MNW:
\eqn\ellCY{w^2 ~=~  z_1^3 ~+~ z_2^6 ~+~ z_3^6  ~-~
{1 \over z_4^6} ~+~ \psi^2 (1+k^2) (z_1 z_2 z_3 z_4)^2 ~+~
k^2 \psi^4 z_1 (z_2 z_3 z_4)^4 \ .}
The moduli are $\psi$ and $k$, and \ellCY\
reduces to \EeightCY\ in the limit $k \to 0$ (one also has to
shift $w$ and re-scale $\psi$).  It is convenient to
think of $k$ as the modulus of a set of Jacobi
elliptic functions, and so we
parametrize $k$ in terms of another variable, $\tau$,
via 
\eqn\kdefn{k ~\equiv~ \vartheta_2^2(0|\tau)/\vartheta_3^2
(0|\tau) \ .}

The holomorphic $3$-form can be represented as
\eqn\holOm{\Omega ~=~  {\psi \over 2 \omega_2} ~
{ z_4~dz_1~dz_2~dz_3\over  w} \ ,}
where  the (constant) normalization factor 
${\psi \over 2 \omega_2}$ will be defined
below.  To isolate the relevant periods one now
follows the approach of \LMW:  Go to a patch with $z_4 =1$,
and make a change of variables $z_1 = \zeta z_2^2$.
Considered as a function of $z_2$, $\Omega$ has
branch cuts.  These cuts disappear when $z_3^6 = 1$.
Integrate $z_2$ around a circle around all these cuts.  Let
$\xi = \psi z_3$, and then the period integrals reduce to:
\eqn\redint{ {1 \over 2 \omega_2} ~ \int 
{d \xi ~ d \zeta \over  \sqrt{\zeta^3 + 1 - (1 + k^2) \zeta^2
\xi^2 + k^2 \zeta \xi^4} } \ .}
There are two types of period integral:
\item{(i)} Integrate $\xi$ between roots of $\xi^6 = \psi^6$
\item{(ii)} Take $\zeta = x \xi^2$ and integrate $\xi$ around a 
circle of large radius.  

Doing the latter first, one is left with a  standard 
elliptic integral in $x$:
\eqn\ellint{{1 \over  \omega_2} ~ \int  { d x  \over  
\sqrt{  (x^3 - (1+k^2) x^2 + k^2 x) } } ~=~  {1 \over 
\omega_2} ~ \int 
{d t  \over  \sqrt{  (1 - t^2)~(1 - k^2 t^2)} } \ ,}
where $x = 1/t^2$.

Shifting $x$, one can recast the elliptic integral in
standard Weierstrass form $\int dx/y$ with 
$y^2 = 4 x^3 - g_2 x - g_3$ and
\eqn\weierform{ g_2 ~=~  \coeff{4}{3}( 1 - k^2 + k^4) \ , 
\qquad   g_3 ~=~ \coeff{4}{27}( 1 + k^2)(2 k^2 -1) (k^2 -2) \ .}
The periods of the torus defined by $x$ and $y$
are denoted by $\omega_1$ and
$\omega_2$.  This defines $\omega_2$ in \holOm, and with this
normalization the  periods \ellint\ are $1$ and $\tau = 
\omega_1/\omega_2$.  The parametrization \kdefn\ gives
the relationship between $k$ and the parameter $\tau$
introduced here.  The parameter $\tau$ is to be identified
with the K\"ahler modulus $t_E$.

Returning to the other period integrals, we first recast them
into a form similar to that of \MNW.  Make a change of variables
$\zeta = {1 \over 4} x/ \xi^{10}$, $ u = - 2 \xi^6$.  The integral 
becomes ${i \over 12 \omega_2} \int dx du /y$ where 
\eqn\newcurve{y^2 ~=~ x^3 ~+~ (1 + k^2) u^2 x^2 + 
k^2 u^4 x ~-~ 2 u^5 \ .}
This can easily be recast into standard Weierstrass form
$y^2 =  4 x^3 - \tilde g_2 x - \tilde g_3$ with:
\eqn\tildeweier{\eqalign{\tilde g_2 ~=~  & \coeff{4}{3}( 1 - k^2 + 
k^4)~ u^4 ~=~  g_2~ u^4\ ,  \cr  \tilde g_3 ~=~ & \coeff{4}{27}
(1 + k^2) (2 k^2 -1) (k^2 -2)~u^6  ~-~ 8~u^5 ~=~ g_3 ~-~ 8~u^5\ .}}
Let  $\tilde \omega_i$ be the periods of this torus, 
and $\tilde \tau = \tilde \omega_1/\tilde \omega_2$.  The
periods we then seek are the indefinite integrals 
$\int du ~\tilde \omega_i/\omega_2$ where $u = -2 \psi^6$. 

Introduce the Eisenstein functions:
\eqn\Eisensteins{\eqalign{E_2(\tau) ~\equiv~ & 1 ~-~ 24 
\sum_{n=1}^\infty ~ \sigma_1(n) ~ q^n \ , \cr
E_4(\tau) ~\equiv~ & 1 ~+~ 240
\sum_{n=1}^\infty ~ \sigma_3(n) ~ q^n \ , \cr
E_6(\tau) ~\equiv~ & 1 ~-~ 504
\sum_{n=1}^\infty ~ \sigma_5(n) ~ q^n \ , }}
where $q = e^{2 \pi i \tau}$ and $\sigma_p(n)$ is
the sum of the $p^{\rm th}$ powers of the divisors
of $n$.  One should recall that $E_4$ and $E_6$ transform
as modular functions of weight $4$ and $6$ respectively,
and that
\eqn\Etwotrf{E_2 \Big({a \tau + b \over c \tau + d}\Big) ~=~ 
(c \tau + d)^2 ~\Big( E_2(\tau) ~+~ {6 \over \pi i}~
{c \over c \tau +d} \Big) \ .}

{}From \tildeweier\ and the standard relationship between
the periods of a torus and the Eisenstein functions one finds:
\eqn\perrelns{ \tilde \omega_2  ~=~ {\omega_2  \over u} ~
\left({ E_4(\tilde \tau ) \over E_4 (\tau) }\right)^{1/4} \ , \qquad
{E_4^3 (\tilde \tau) \over E_6^2(\tilde \tau)} ~=~ {E_4^3 (\tau) \over
(E_6(\tau)- v)^2} \ ,}
where $v = {1 \over 27} ({\pi \over \omega_2})^6 {1 \over u}$.
The second equation in \perrelns\ may be used to determine $\tilde 
\tau$ as a function of $\tau$ and $u$.  This can then be substituted
into the first equation to give $\tilde \omega_2/\omega_2$ in terms
of $\tau$ and $u$.  One then gets the other period integral from 
$\tilde \omega_1/\omega_2 = \tilde \tau \tilde \omega_2/\omega_2$.

Therefore, the other two periods of $\Omega$ are:
\eqn\phidefns{\varphi ~\equiv~  {1 \over 2 \pi i }~\int ~ 
{dv \over v} ~ \left({ E_4(\tilde \tau ) \over E_4 (\tau) }
\right)^{1/4} \ ;  \qquad \varphi_D ~\equiv~  {1 \over 2 \pi i }~
\int ~ {dv \over v} ~ \tilde \tau ~
\left({ E_4(\tilde \tau ) \over E_4 (\tau) }\right)^{1/4} \ ,}
where $\tau$ is fixed, and $\tilde \tau$ is obtained from \perrelns.  

\newsec{Curve counting and the instanton expansion}

To count the BPS states we want the Taylor expansion of 
the periods \phidefns\ about $v = 0$.  To evolve these series it is
useful to recall that:
\eqn\Eisenderivs{\eqalign{E_2  ~=~ & q {d \over d q} \log 
(\Delta) \ ; \qquad 
q {d \over d q} E_2  ~=~  \coeff{1}{12}~\big( E_2^2 ~-~
E_4 \big) \ ; \cr
q {d \over d q} E_4  ~=~ & \coeff{1}{3}~\big( E_2 E_4 ~-~
E_6\big) \ ; \qquad 
q {d \over d q} E_6  ~=~  \coeff{1}{2}~\big( E_2 E_6 ~-~
E_4^2 \big)   \ ,}}
where $\Delta = \eta^{24} = q \prod_{n=1}^\infty (1 - q^n)^{24}
= {1 \over 1728}(E_4^3 - E_6^2)$.  Also 
recall that the modular invariant $j$ is given by 
$j = E_4^3/\Delta$.  From \perrelns\ one obtains:
\eqn\tauexp{\tilde \tau ~=~ \tau ~+~ \sum_{j=1}^\infty ~a_j~ v^j 
\ ,}
and the $a_j$ can be obtained using \Eisenderivs.  The first few
terms are:
\eqn\acoeffs{\eqalign{ a_1 ~=~ & 2 E_4/\Delta \ , \qquad a_2 ~=~ 
-{1 \over 3  \Delta^2}~E_4~\big(5 E_6 ~+~ E_2 E_4 \big)  \ , \cr
a_3 ~=~ & -{1 \over  \Delta^3}~E_4~\big(\coeff{31}{54}~E_4^3 ~+~
\coeff{40}{27}~E_6^2 ~+~ \coeff{5}{9}~E_2 E_4 E_6  ~+~ 
\coeff{1}{18}~ E_2^2 E_4^2 \big)  \ .}}
Note that the second equation in \perrelns\ is invariant under
the combined modular transformations:
\eqn\modtrf{\tau ~\to~ {a \tau + b \over c \tau + d} \ , \qquad
\tilde \tau ~\to~ {a \tilde \tau + b \over c \tilde \tau + d} \ , 
\qquad v ~\to~ (c \tau + d)^6 ~ v \ .}
The fact that one must simultaneously transform $\tau$ and
$\tilde \tau$ follows from requiring an expansion of the form
\tauexp.  It is also easy to verify that the function 
\eqn\nicecomb{\varphi_D - \tau \varphi ~=~  {1 \over 2 \pi i }~
\int ~ {dv \over v}~  (\tilde \tau - \tau)~ \left({ E_4(\tilde \tau ) 
\over E_4 (\tau) } \right)^{1/4} ~\equiv~ \sum_{j=1}^\infty ~
b_j~v^j \ ,}
considered as a function of $\tau$ and $v$, transforms 
as  a modular function of weight $-2$ under \modtrf.  Thus upon
substituting \tauexp\ into \nicecomb, and expanding in a 
power series is $v$, the coefficient, $b_j$, of $v^j$ must be
an exactly modular function of weight $-2 - 6j$.  One also
sees from \perrelns\ that the only pole in $\varphi_D -
\tau \varphi$ is at $\tau = i \infty$, and from the examining
the form of $a_j$, one sees that the $b_j$ must be of the
form $P_{6j-2}(E_4,E_6)/\Delta^j$, where $P_{6j-2}(E_4,E_6)$
is a polynomial in $E_4$ and $E_6$ alone, such that it
has modular weight $6j-2$.

Up to ``constants'' of integration, $\delta(\tau)$ and 
$\delta_D(\tau)$, (which could be arbitrary
functions of $\tau$), the periods $\varphi$ and 
$\varphi_D$ are $t_S$ and $\del {\cal F}/\del t_S$:
\eqn\partF{ t_S ~=~ \varphi ~-~ \delta(\tau) \ ;
\qquad  {\del {\cal F} \over \del t_S} ~=~ \varphi_D ~-~ 
\delta_D(\tau) \ .}
Let $q_S = e^{2 \pi i t_S}$ and recall $q  =  e^{2 \pi i \tau} = 
e^{2 \pi i t_E}$.  Define $C_{t_S t_S t_S} = {\del^3
{\cal F} \over \del t_S^3}$, and recall that it has an 
instanton expansion:
\eqn\instexp{C_{t_S t_S t_S} ~=~ \sum_{n_1,n_2}~
(-1)^{n_1+1}N_{n_1,n_2}~{n_1^3 ~ q_S^{n_1} q^{n_2} \over 1 ~-~ q_S^{n_1} 
q^{n_2}} \ , }
where $N_{n_1,n_2}$ is the number of rational curves with
$d_D = n_1$ and $d_E = n_1 + n_2$.

To develop the instanton expansion we therefore have to substitute
\tauexp\ into the expression for $\varphi$, expand the series in
$v$.  One then inverts this series to obtain a series for $v$ in 
powers of $e^{2 \pi i \varphi}$, with coefficients that are 
functions of $\tau$.  This is then substituted into the series
expansion for $C_{t_S t_S t_S}$.   The result is a series of
the form:
\eqn\insttau{C_{t_S t_S t_S} ~=~ \sum_{n=1}^\infty~
 \widetilde F_n(\tau)~e^{ 2 \pi i n \varphi} \ ~=~ 
\sum_{n=1}^\infty~ F_n(\tau)~q_S^n \ , }
where $F_n(\tau) = \widetilde F_n(\tau)~e^{2 \pi i n \delta(\tau)}$.
It is trivial to see that $\widetilde F_1(\tau) = a_1(\tau)$, 
where $a_1$ is given in  \acoeffs. The first term in this series
\insttau\ was computed in \KMV, and is given by:
\eqn\Etheta{F_1(\tau) ~=~ {E_4 \over q^{1/2}~\eta^{12}} \ .}
(The factor in the denominator is $q^{1/2}$ and not $q^{-1/2}$,
since this is the coefficient of $q_S = q e^{2 \pi i t_D}$.) 
This fixes $\delta(\tau) = {1 \over 2 \pi i }~\log( -{1 \over 2} 
q^{-1/2} \eta^{12})$.

Using {\it Mathematica$^{TM}$}, one can easily compute the $F_n$
to fairly high order ($n=12$).  Defining 
$G_n = q^{n/2} F_n$, one finds that $G_n$ has the
form of 
\eqn\GQeq{
G_n=Q_{6n-2}(E_2,E_4,E_6)/\Delta^{n/2} 
}
where $Q_{6n-2}$ is a polynomial of degree $6n-2$ (where
$E_2,E_4$ and $E_6$ are given weights $2,4$ and $6$ respectively).
Because of the presence of $E_2$ and the half powers of
$\Delta$, the functions $G_n$ are
not quite modular functions of weight $-2$.
The first few $G_n$ are:
\eqn\Gresult{\eqalign{G_1 ~=~ & {E_4 \over\Delta^{1/2}} \ , \qquad
G_2 ~=~  { E_4 \, \left( E_2\,E_4 ~+~  2\, E_6 \right)  \over 3\,
\Delta } \ , \cr
G_3 ~=~ & {1 \over 576\, \Delta^{ 3 \over 2}}{E_4\,\left( 54\,
{{E_2}^2}\, {{E_4}^2} ~+~ 109\,{{E_4}^3} ~+~  216\,E_2\,E_4\,E_6 ~+~ 
197\,{{E_6}^2} \right) } \ , \cr 
G_4 ~=~ & {1 \over 972\, \Delta^2} E_4\, \Big( 24\, E_2^3 \, 
{{E_4}^3} ~+~  109\, E_2\, {{E_4}^4} ~+~  144\,{E_2^2}\, E_4^2 \, 
E_6 ~+~  272\,{{E_4}^3} \, E_6 ~+~ \cr  & 
\qquad \qquad \quad 269\, E_2\, E_4\, {{E_6}^2} ~+~ 
154\,{{E_6}^3} \Big) \cr
G_5 ~=~ & {1 \over 2985984\, \Delta^{5 \over 2} }  E_4\,\Big(  18750\,
{{E_2}^4}\, {{E_4}^4} ~+~  136250\,{{E_2}^2}\,{{E_4}^5} ~+~  
116769\,{{E_4}^6} ~+~ \cr & \qquad \qquad \qquad 150000\,{{E_2}^3}\,
{{E_4}^3}\,   E_6 ~+~   653000\,E_2\,{{E_4}^4}\, E_6 ~+~  426250\,
{{E_2}^2}\,{{E_4}^2}\,  {{E_6}^2} ~+~ \cr & \qquad \qquad \qquad
772460\,{{E_4}^3}\,{{E_6}^2} ~+~  
 505000\,E_2\,E_4\,  {{E_6}^3} ~+~ 207505\,{{E_6}^4} \Big) \cr
G_6 ~=~ & {1 \over 74649600\, \Delta^3} E_4\,\Big( 116640\, E_2^5 \,
 E_4^5 ~+~ 1177200\,E_2^3 \, E_4^6  ~+~  2398867\,E_2\,{{E_4}^7} ~+~ 
\cr & \qquad \qquad \qquad 1166400\, E_2^4\, E_4^4\, E_6 ~+~ 8229600\, 
E_2^2  \,E_4^5\, E_6 ~+~  6703718\,{{E_4}^6}\,E_6 ~+~ \cr & \qquad \qquad 
\qquad 4460400\,{{E_2}^3}\, {{E_4}^3}\,{{E_6}^2} ~+~ 18894730\,E_2\,
{{E_4}^4}\,   {{E_6}^2} ~+~ \cr & \qquad \qquad \qquad 8100000\,{{E_2}^2}\,
{{E_4}^2}\,  E_6^3 ~+~  14280020\,{{E_4}^3}\, {{E_6}^3} ~+~  
6922915\,E_2\,E_4\,  {{E_6}^4} ~+~  \cr & \qquad \qquad \qquad 2199110\,
{{E_6}^5} \Big)  \ . } }
These functions are completely consistent with the numbers
generated in \KMV.

The first function in \Gresult\ has the simple interpretation 
as the $E_8$ root lattice partition function multiplied by
the partition function of four bosonic oscillators coming
from the space-time \KMV.  The second function is almost as 
simple:  it can be rewritten as $G_1 G_1'$, where $G_1' =
q {q \over d q} G_1$.  Since this partition function
represents momentum excitations of a doubly wound string, 
one naively expects a tensor product, or $G_1^2$.  However,
since it is a bound state, it cannot be a simple tensor
product.  The derivative with respect to $\tau$ pulls down
a hamiltonian and thus removes one of the two independent 
translation invariances of the doubly wound state.  Unfortunately 
the higher $G_n$ do not appear to admit such a simple interpretation.
However, as we will see in the next section, the $G_n$ are
indeed related to one another.
 
\newsec{Modular properties and a recurrence relation}

There is no {\it a priori} reason to expect the 
non-critical string to exhibit $T$-duality, but the $G_n$
are almost modular functions (of weight $-2$), and so the 
spectrum of BPS states is almost invariant under $\tau \to 
{a \tau + b \over c \tau + d}$.  There are three places in
which the BPS spectrum fails to be modular invariant:
(i) the bare factors of $q$ in $F_n = q^{n/2} G_n$, 
(ii) the odd powers of $\Delta^{1/2}$ in $G_n$, and
(ii) the anomalous modular behaviour of $E_2$. The first
problem has a trivial cure: one simply redefines $t_S$.
The second problem means that one really has only a subgroup
of the modular groups as a symmetry.  Alternatively, one
can work with the full modular group, and remember to make
appropriate changes of sign in the odd powers of $\Delta^{1/2}$ .
The anomalous behaviour of $E_2$ also has a cure, but at the cost 
of holomorphy. That is, the function
\eqn\Etwohat{\widehat E_2 (\tau)  ~=~ E_2(\tau) ~-~ 
{3 \over \pi}~ {1 \over Im(\tau)} \ , }
satisfies
\eqn\Etwotrf{\widehat E_2 \Big({a \tau + b \over c \tau + d}
\Big) ~=~  (c \tau + d)^2  \widehat E_2(\tau)  \ .}
Define $\widehat {\cal G}_n$ to be $G_n/n^3$ but with $E_2$ 
replaced by $\widehat E_2$, and introduce:
\eqn\hatGdefn{\widehat {\cal G}(\sigma,\tau) ~=~ 
\sum_{n=1}^\infty~ \widehat {\cal G}_n(\tau)~ e^{2 \pi i n 
\sigma} \ .}
One can recover the instanton expansion from this by taking
three derivatives with respect to $\sigma$, setting
$\sigma = t_S + {1 \over 2} \tau$, and by sending $\bar \tau
\to \infty$ while holding $\tau$  fixed.  The function 
$\widehat {\cal G}$ is also modular invariant.  To this extent
the non-critical string exhibits a $T$-duality.

The function $\widehat {\cal G}$ exhibits another remarkable 
property: it satisfies a recurrence relation that is reminiscent
of the holomorphic anomaly equation in Kodaira-Spencer 
theory \BCOV.  This follows from a ``modular anomaly'' recurrence
relation that is satisfied by the $F_n$.  
Let $f_n = F_n/n^3$, and view it as a function
of the variables $E_2, E_4$ and $E_6$, then we will show that
\eqn\recursion{ {\partial f_n \over \partial E_2} ~=~ {1\over24}~
\sum_{m=1}^{n-1}~ m(n-m)~f_mf_{n-m} \ .}
If one replaces $E_2$ by $\widehat E_2$, one has  $ {\partial \over
\partial \bar \tau} = - {3i \over 2 \pi} {1 \over (Im(\tau))^2} 
{\partial  \over \partial \widehat E_2}$.  Hence it follows that
$\widehat {\cal G}$ satisfies:
\eqn\Gdiffeq{{\partial \widehat {\cal G} \over \partial \bar 
\tau} ~=~ - {i \over 16 \pi} {1 \over (Im(\tau))^2}~~{\partial 
\widehat {\cal G} \over \partial \sigma} ~~{\partial 
\widehat {\cal G} \over \partial \sigma} \ .}

This equation is almost exactly of the form of the holomorphic
anomaly equation of \BCOV, although the physical origins of
the problem are rather different.  In \BCOV\ the holomorphic
anomaly was used to relate partition functions on higher genus Riemann 
surfaces to those of lower genus.  In our case, the anomaly equations 
are used  to relate multi-instanton expansions to lower instanton terms.  
This probably explains one key difference between the anomaly equations.  
In \BCOV, the anomaly equation for $F_g$ where $g$ is the genus, contains 
the piece $\partial^2F_{g-1}$.  Such a term arises from pinching off a 
handle on the Riemann surface.  It is hard to imagine an analogous 
process for a multi-instanton.  

Still, the similarities are striking.  The instanton expansion consists
of maps onto a target space that is a torus with modulus $\tau$.  
The factor of $1/(Im(\tau))^2$ in \Gdiffeq\ has the interpretation
of a metric $g^{\sigma\sigma}$ on the torus \BCOV, while the fact that the 
differential equation involves a derivative with respect to 
$\bar \tau$ on the left and derivatives with respect to 
$\sigma$ on the right is consistent with the form of the
classical intersection form:  Since $\varphi_D = \tau \varphi$
it follows that $C_{\tau t_S t_S} = 1$.  There is no corresponding equation
for $\partial_{\bar\phi}\widehat{\cal G}$ since the classical
(non-instanton) part of $C_{t_S t_S t_S}$ is zero.  Given
this close correspondance, it is tempting to conjecture that
$e^{ 2 \pi i t_S}$ might be a loop expansion parameter for
the non-critical string.


The recurrence relation in \recursion\ turns out to be a powerful tool
in computing the $G_n$ of the previous section.  Given the lower $G_m$,
the recurrence relation determines 
$G_n$ ($n>m$) up to a piece $E_4K_n/\Delta^{n/2}$, where $K_n$
is a modular form of weight $6(n-1)$.  The space of such forms has  
dimension $[(n+1)/2]$, thus to completely determine $G_n$, we need
to compute $[(n+1)/2]$ coefficients.  We actually have more than enough
information to do this, since we know that the $q$ expansion of $G_n$
has the form $G_n=q^{-n/2} +{\rm O}(q^{n/2})$\foot{This follows from 
\instexp\  and the fact that $n_1\ge1$ $n_2\ge0$ and $N_{1,0}=1$}.  
Hence, all we need to do
is adjust the coefficients in $K_n$, such that leading term in $G_n$ is
$q^{-n/2}$ and the rest of the negative powers in $G_n$
have coefficients that are zero.  A useful check is that the nonnegative 
powers up to the $q^{n/2}$ term also have zeros for coefficients.
Using {\it Mathematica$^{TM}$}, we can easily generate the $K_n$,
the first 12 of which are given by

$$\eqalign{K_1=&1,\qquad K_2={2E_6\over3},\qquad K_3=
{1\over576}\left({109\,{{{E_4}}^3} + 197\,{{{E_6}}^2}}
   \right),\qquad K_4=
{{{E_6\over{486}}\,\left( 136\,{{{E_4}}^3} + 
       77\,{{{E_6}}^2} \right) }},\cr
K_5=&{1\over{2985984}}\left({{116769\,{{{E_4}}^6} + 
     772460\,{{{E_4}}^3}\,{{{E_6}}^2} + 
     207505\,{{{E_6}}^4}}}\right),\cr
K_6=&{{{E_6}\over{37324800}}\,\left( 3351859\,{{{E_4}}^6} + 
       7140010\,{{{E_4}}^3}\,{{{E_6}}^2} + 
       1099555\,{{{E_6}}^4} \right) },\cr
K_7=&{1\over{386983526400}}\Bigl(3214033725\,{{{E_4}}^9} + 
     46377519701\,{{{E_4}}^6}\,{{{E_6}}^2} + 
     47881472765\,{{{E_4}}^3}\,{{{E_6}}^4}\cr
&\qquad + 
     4721253845\,{{{E_6}}^6}\Bigr),\cr
K_8=&{{E_6}\over{111106598400}}\,\Bigl( 2874313704\,
        {{{E_4}}^9} + 
       13496101157\,{{{E_4}}^6}\,
        {{{E_6}}^2} + 
       8126197310\,{{{E_4}}^3}\,{{{E_6}}^4}\cr
&\qquad + 
       551920565\,{{{E_6}}^6} \Bigr),
}$$
\eqn\KeqII{\eqalign{
K_9=&{1\over{1213580338790400}}\Bigl(2168558256025\,{{{E_4}}^{12}} + 
     54762568177568\,{{{E_4}}^9}\,
      {{{E_6}}^2}\cr
&\qquad + 
     125727877850316\,{{{E_4}}^6}\,
      {{{E_6}}^4} + 
     49166052530000\,{{{E_4}}^3}\,
      {{{E_6}}^6} + 2423088666145\,{{{E_6}}^8}
     \Bigr),\cr
K_{10}=&{{E_6}\over 88470006697820160}\,\Bigl( 621851537315031\,
        {{{E_4}}^{12}} + 
       5138509650200980\,{{{E_4}}^9}\,
        {{{E_6}}^2}\cr
&\qquad + 
       6932453167897530\,{{{E_4}}^6}\,
        {{{E_6}}^4} + 
       1889989579331700\,{{{E_4}}^3}\,
        {{{E_6}}^6} + 
       70283214345575\,{{{E_6}}^8} \Bigr),\cr
K_{11}=&{1\over{4246560321495367680000}}\Bigl(
1644909843291474375\,{{{E_4}}^{15}} + 
     64293839773877897511\,{{{E_4}}^{12}}\,
      {{{E_6}}^2}\cr
&\qquad + 
     261044867981347260580\,{{{E_4}}^9}\,
      {{{E_6}}^4} + 
     230239896247913645940\,{{{E_4}}^6}\,
      {{{E_6}}^6} \cr
&\qquad+ 
     46023695034064491975\,{{{E_4}}^3}\,
      {{{E_6}}^8} + 
     1331341121000896775\,{{{E_6}}^{10}}\Bigr),\cr
K_{12}=&{E_6\over
5352435405218119680000}\,\Bigl(9853164552615074200\,
        {E_4^{15}} + 
       126458286011220239911\,{{{E_4}}^{12}}\,
        {{{E_6}}^2} \cr
&+
       303024347024677902580\,{{{E_4}}^9}\,
        {{{E_6}}^4} + 
       187415787390550146890\,{{{E_4}}^6}\,
        {{{E_6}}^6} \cr
&+ 
       28525664977566703100\,{{{E_4}}^3}\,
        {{{E_6}}^8} + 
       657760456052320775\,{{{E_6}}^{10}} \Bigr) }
}
The coefficients of these polynomials  are always
positive.  At this time, we know of no other system where these
particular modular forms appear naturally.

We conclude by proving the recurrence relation \recursion.
This relation is equivalent to:
\eqn\deqphi{
{\partial \over \partial E_2}~(\varphi_D - \tau \varphi) ~=~ 
{2\pi i \over 24}~ {\partial \over \partial \varphi}~(\varphi_D - 
\tau  \varphi)^2 \ ,}
where $\varphi_D - \tau \varphi$ is viewed as a function of
$\varphi$, $E_2,E_4$ and $E_6$.  Now recall that the coefficients
$b_j$ in \nicecomb\ are independent of $E_2$,  and so
$\varphi_D - \tau \varphi$ inherits its $E_2$ dependence
{\it only} through the implicit dependence of $v$ on 
$\varphi$, $E_2,E_4$ and $E_6$.  As a result, \deqphi, and
hence \recursion, is equivalent to 
\eqn\vandphi{ {\partial v \over \partial E_2}  ~=~ 
{2 \pi i \over 12}~ (\varphi_D - \tau  \varphi) ~
{\partial v \over \partial \varphi} \qquad {\rm or}
\qquad (\varphi_D - \tau  \varphi)  ~=~ {12\over 2\pi i }~
{ \partial_{E_2} v  \over \partial_\varphi v  } \ .}

Differentiating the first equation in \phidefns\ with respect to
both $\varphi$ and $E_2$ yields
\eqn\phiveq{\eqalign{ 2\pi i ~=~ & {\partial_\varphi v \over v}
\bigg({E_4(\tilde \tau) \over E_4(\tau)}\bigg)^{1/4} \ , \cr
0 ~=~ & {\partial_{E_2}  v \over v} \bigg({E_4(\tilde \tau) \over 
E_4(\tau)}\bigg)^{1/4} ~+~ \int^v ~{dv\over v} ~\partial_{E_2} 
\bigg ({E_4(\tilde \tau)\over E_4(\tau)}\bigg)^{1/4} \ ,}}
where the $E_2$ derivative acting on the integrand is
for the explicit $E_2$ dependence ({\it i.e.} not the implicit dependence
in $v$).  From these two equations
one obtains:
\eqn\phivEeq{
{\partial_{E_2} v \over\partial_\varphi v} ~=~ ~-~{1\over2\pi i}~
\int^v~ {dv\over v} ~\partial_{E_2} \bigg({E_4(\tilde \tau) \over 
E_4(\tau)}\bigg)^{1/4}.}
Using this in \vandphi\ one sees that the recurrence relation
is equivalent to:
\eqn\inteq{
\int^v~ {dv\over v} ~ \bigg ({E_4(\tilde \tau) \over E_4(\tau)}
\bigg)^{1/4} ~(\tilde \tau - \tau) ~=~ -{ 12 \over 2\pi i}~\int^v~
{dv \over v}~\partial_{E_2} \bigg ({E_4(\tilde \tau) \over 
E_4(\tau)}\bigg)^{1/4} \ ,}
or equating integrands, one has
\eqn\integrands{ \bigg ({E_4(\tilde \tau) \over E_4(\tau)}
\bigg)^{1/4} ~(\tilde \tau - \tau) ~=~ -{ 12 \over 2\pi i}~
\partial_{E_2} \bigg ({E_4(\tilde \tau) \over 
E_4(\tau)}\bigg)^{1/4} \ .}
Since $\Bigl({E_4(\tilde \tau) \over E_4(\tau)}
\Bigr)^{1/4} ~(\tilde \tau - \tau)$ is
a modular function of weight $-2$, it cannot have any explicit 
$E_2$ dependence.  Using this fact in the right hand side of 
\integrands\ means that \integrands\ is equivalent to:
\eqn\tauEeq{ \partial_{E_2} (\tilde \tau - \tau) ~=~ -{2\pi i\over12} 
(\tilde \tau - \tau)^2 \ .}
Thus proving \tauEeq\ is equivalent to establishing the recurrence
relation.  Let $H = \tilde \tau - \tau$, then under the modular
transformation \modtrf, 
\eqn\Hmodtrf{ H ~=~ (\tilde \tau - \tau) ~\to~ {(\tilde \tau - \tau)
\over (c \tau +d) (c \tilde \tau + d)} ~=~ {H /(c \tau + d)^2 
\over 1 + c H/ (c\tau + d)} \ .}
Let $h$ be defined by 
\eqn\hdefn{  H ~=~ {h \over1 ~+~ {2\pi i\over12} E_2 h} \ .}
Then \Hmodtrf\ and \Etwotrf\ implies that $h \to 
(c\tau +d)^{-2} h$, {\it i.e.} it is a modular function
of weight $-2$.  It therefore cannot have any {\it explicit}
$E_2$ dependence, and so from \hdefn\ one has:
\eqn\finally{ \partial_{E_2} H ~=~ ~-~ {2\pi i\over12}~
{h^2 \over \big(1 ~+~ {2\pi i\over12} E_2 h \big)^2} ~=~
- {2\pi i\over12}~ H^2 \ .}
This establishes \tauEeq, and hence proves the recurrence
relation.

\newsec{Asymptotic degeneracies of BPS states}

An important question is the asymptotic behaviour for the
degeneracies of BPS states, $N_{n_1,n_2}$, as $n_1$ and $n_2$ approach 
large values in \instexp.  By knowing the behaviour of
the degeneracies one can compare the entropy with that of other 
interesting physical systems.

The contributions of multicoverings becomes
negligible in the asymptotic limit, so to a very good approximation
\eqn\NFrel{N_{n_1,n_2} ~=~ (-1)^{n_1+1}F_{n_1,n_2}/n_1^3}
where $F_{n_1,n_2} = G_{n_1,n_2+n_1/2}$ is the coefficient of 
$q^{n_2}$ in $F_{n_1}$.  The coefficient $G_{n,m}$ is given by
\eqn\Gcoeff{ G_{n,m }~=~ {1\over2\pi i}\oint {dw\over w^{m+1}}~
G_n (\tau) \ ,}
where $w = e^{2 \pi i \tau}$, and the integral is taken around
any circle centered on the origin in the $w$-plane.
We now follow the method in \GSW\ of making a saddle
point approximation of this integral for large values of $m$.
To do this we first recall the normalization
requirement that $G_n \sim 1. q^{-n/2}$ as $q \to 0$.
To get a well-behaved and consistent approximation, one
first makes a modular inversion, and considers the limit
in which $\tau $ is small.  Recall that $E_4$ and
$E_6$ are modular forms of weight $4$ and $6$, 
$E_2$ transforms as in \Etwotrf, and that $\Delta^{1/2}(-1/\tau) =-
\tau^6\Delta^{1/2}(\tau)$.   If one keeps only the leading terms in
$e^{-2 \pi i/\tau}$, one can neglect the anomalous modular pieces
of the transformation of $E_2$, and so with the exception of
the sign in the the transformation of $\Delta^{1/2}$, $G_n$
to leading order behaves like a modular function of weight $-2$.
Keeping only the leading term in the $q$-expansion of $G_n$
then yields:
\eqn\Gcoeffapp{ G_{n,m} ~\approx~{1\over2\pi i} \oint {dw\over w^{m+1}}~
(-1)^n \left(-{2\pi i\over\log w}\right)^2 ~ \exp\left({-4\pi^2 n
\over 2\log w}\right) \ .}
The saddle point is at $\tau = i \sqrt{n \over 2m }$, which is
indeed small as $m$ becomes large.  Continuing in the usual way
with the standard saddle point approximation, we find
\eqn\Gsad{G_{n,m} ~\approx~ -(-1)^n ~{n^{5/4}\over(2m)^{7/4}}~
\exp\left(2\pi\sqrt{2nm}\right). }
Hence, the asymptotic approximation for $N_{n_1,n_2}$ is
\eqn\Nasymp{ N_{n_1,n_2}~\approx~ (n_1^2+2n_1n_2)^{-7/4}~
\exp\left(2\pi\sqrt{n_1^2+2n_1n_2}\right),
}
which is valid for both $n_1$ and $n_2$ large.

We can also consider the limit where $n_1$ is large and $n_2$ is zero.  
{}From  \Nasymp, the naive behaviour is 
\eqn\naive{N_{n_1,0} ~\sim~ (n_1)^{-7/2}\exp(2\pi n_1) \ .}
Let us compare this with the
exact asymptotic behaviour.  We can project onto the $n_2=0$ states by
letting $\tau\to i\infty$.   This reduces the model to the case
considered in \LMW\ and \MNW.  Now the instanton expansion is given by
\eqn\instanton{ \varphi_D=-\half\varphi^2+\half\varphi+ {5 \over 12}
-\sum_{n=1} {C_n\over 4\pi^2n^2}~ e^{2\pi in\varphi},
\qquad\qquad \partial^3_\varphi\varphi_D=\sum_{n=1}2\pi i n~C_n~
e^{2\pi i n\varphi} \ .}
For the asymptotic expansion we can again ignore the multicoverings.  Thus
the degeneracy $d_n$ is
\eqn\dn{d_n ~\approx~ -(-1)^n ~C_n/n^3 \ .}

The instanton expansion diverges at the conifold point $u=1$.  Near this
point $\varphi$ and $\varphi_D$ are approximately \LMW:
\eqn\phieq{\eqalign{
\varphi& ~=~ c ~-~ {i\over 2\pi}(\phi_D-1)\log(u-1)~+~...\cr
\varphi_D& ~=~ 1~+~{1\over2\pi}(u-1) \ .}}
So to a good approximation we have
\eqn\phiDeq{\eqalign{
\varphi_D& ~=~ 1 ~+~ 2\pi i(\varphi-c)/\log(\varphi-c) ~+~...\cr
{\partial_\varphi}^3\varphi_D& ~=~ {2\pi i\over((\varphi-c)
\log(\varphi-c))^2}~+~...}}
Hence, when $\varphi$ is close to $c$, we have
\eqn\dneq{ \sum_n~ 2\pi i~ n^4~ (-1)^{n+1}~d_n ~e^{2\pi i n\varphi}~
\approx~ \int dn~2\pi i ~n^4 ~(-1)^{n+1}~d_n~ e^{2\pi in\varphi} ~
\approx~ {2\pi i \over((\varphi-c)\log(\varphi-c))^2 }\ .}
If we use the ansatz that $d_n \sim n^a(\log n)^b e^{-2\pi i c n}$, then
a simple saddle point calculation yields
\eqn\dnasym{
d_n ~\approx~ (2\pi)^{3/2}n^{-3}(\log n)^{-2}(-1)^ne^{-2\pi i cn+1}.}
To find $c$ note that we have to match the behaviour of $\varphi$ and
$\varphi_D$ at the conifold point with their behaviour at the $E_8$ point 
$u=0$. The behaviour is matched by noting that \LMW
\eqn\phihyper{\eqalign{
\varphi_D& ~=~ \kappa\int {du\over u}\left(\xi F_0(u)+{1\over\xi}
F_1(u)\right) \cr  \varphi& ~=~ \kappa\int {du\over u}\left(e^{2\pi i/3}
\xi F_0(u)+{e^{-2\pi i/3}\over\xi} F_1(u)\right) \ ,}}
where $\kappa=i~3^{1/4}/(4\pi^{3/2})$ and 
$\xi=-i~ 3^{1/4}\Gamma(1/3)^3/(2^{2/3}\pi^{3/2})$ and 
\eqn\hypereq{
F_0(u)~=~u^{1/6}~{}_2F_1({1\over6},{1\over6};{1\over3};u)\qquad\qquad
F_1(u)~=~u^{5/6}~{}_2F_1({5\over6},{5\over6};{5\over3};u).}
To find the integration constants,  we know that as $u\to 0$, 
$\varphi\to 0$ and  $\varphi_D\to 0$.  Hence, the integrals in 
\phihyper\ are integrated from $u=0$ to $u$.  Therefore, $c$ is given by
\eqn\ceq{\eqalign{c& ~=~ \kappa\int_0^1 {du\over u} \left(e^{2\pi i/3}
\xi F_0(u)+{e^{-2\pi i/3}\over\xi}  F_1(u)\right)\cr & ~=~ 
-\half ~+~ i\left(\left({9\Gamma(1/3)^3\over2^{5/3}\pi^3}\right)
{}_3F_2({1\over6}, {1\over6},{1\over6};{1\over3},{7\over6};1) ~-~ 
{\sqrt{3}\over2}\right) \ ,}}
where ${}_3F_2$ is a generalized hypergeometric function evaluated 
at $u=1$.  Hence we find that
\eqn\dnapprox{ d_n ~\approx~ (2\pi)^{3/2}n^{-3}(\log n)^{-2}
\exp\left((2\pi-\alpha)n-1\right) \ .}
where $\alpha$ is found from \ceq\ and is approximately $\alpha=.451967$.
Comparing this with \naive, we see that
the naive result is not such a terrible approximation.

\goodbreak
\vskip2.cm\centerline{\bf Acknowledgements}
\noindent
We would like to thank W.~Lerche and P.~Mayr for
valuable discussions.  N.W.~is also grateful to the ITP in
Santa Barbara, and the Institute for Advanced Study in
Princeton for hospitality while this work was being done.
This work is supported in part
by funds provided by the DOE under grant number DE-FG03-84ER-40168,
and by the National Science Foundation under grant No. PHY94-07194.

\goodbreak
\listrefs

\vfill
\eject
\end